# Formation of BCR Oligomers Provides a Mechanism for B cell Affinity Discrimination

Philippos K. Tsourkas[†], Somkanya C. Das[†§], Paul Yu-Yang[†§], Wanli Liu[‡], Susan K. Pierce[‡], and Subhadip Raychaudhuri[*†]

[†]*Dept. of Biomedical Engineering*
*University of California, Davis*
*One Shields Avenue, Davis, CA 95616, USA*

[‡]*Laboratory of Immunogenetics,*
*National Institute of Allergy and Infectious Diseases,*
*National Institutes of Health, Rockville, MD 20852*

[§] Authors contributed equally

[*]Address correspondence (including requests for supplemental figures) to:
Subhadip Raychaudhuri, Dept. of Biomedical Engineering, University of California, One Shields Avenue, Davis, CA 95616, Tel: (530) 754-6716
raychaudhuri@ucdavis.edu

## ABSTRACT

B cells encounter antigen over a wide affinity range, from $K_A=10^5$ $M^{-1}$ to $K_A=10^{10}$ $M^{-1}$. The strength of B cell signaling in response to antigen increases with affinity, a process known as "affinity discrimination". In this work, we use a computational simulation of B cell surface dynamics and signaling to show that affinity discrimination can arise from the formation of BCR oligomers. It is known that BCRs form oligomers upon encountering antigen, and that the size and rate of formation of these oligomers increase with affinity. In our simulation, we have introduced a requirement that only BCR-antigen complexes that are part of an oligomer can engage cytoplasmic signaling molecules such as Src-family kinases. Our simulation shows that as affinity increases, not only does the number of collected antigen increases, but so does signaling activity. Our results are also consistent with the existence of an experimentally-observed threshold affinity of activation at $K_A=10^5$-$10^6$ $M^{-1}$ (no signaling activity below this affinity value) and affinity discrimination ceiling of $K_A=10^{10}$ $M^{-1}$ (no affinity discrimination above this affinity value). Comparison with experiments shows that the time scale of dimer formation predicted by our model (less than 10 s) is well within the time scale of experimentally observed association of BCR with Src-family kinases (10-20 s).

## INTRODUCTION

B cells, the cells responsible for antibody production, are activated by recognizing antigen (Ag) through the B cell antigen receptor (BCR) located on their surface. The strength of BCR signaling in response to stimulation by antigen is known to increase monotonically with antigen affinity, a phenomenon known as "affinity discrimination" [1-9]. B cell affinity discrimination is critical to the process of affinity maturation that results in the production of high affinity antibodies [9], and is thus important in applications such as vaccine design [9]. The precise mechanisms by which B cells receptors sense antigen affinity are still not fully known [10]. While the first studies of B cell affinity discrimination focused on multivalent antigen encountered in soluble form, recent research shows that antigens presented on the surface of antigen presenting cells (APCs), typically dendritic cells or macrophages, are potent stimulators of B cells [4,10-21].

Further studies have shown that during the initial stages of contact between B cells and APCs, micro-clusters of 10-100 BCR-antigen complexes form on protrusions of the B cell surface [8,9,13,22,23]. These micro-clusters are signaling-active [8,9,13,22,23], as they trigger affinity-dependent spreading of the B cell surface over the antigen presenting cell surface, increasing the cell-cell contact area [8]. This spreading response leads to the formation of micro-clusters at the leading edges [8,22], culminating in the formation of the immunological synapse [7,8,10,12]. It has also been shown that early signaling events (~100 seconds) such as $Ca^{2+}$ flux, as well as antigen accumulation in the immunological synapse, all increase with antigen affinity [8,9]. Thus, affinity discrimination has been observed at the earliest stages of contact between B cell receptors and antigen [8,9].

However, little is known about how B cells discriminate between membrane antigens of varying affinity at the level of BCR-antigen micro-clusters. In previous work, we showed that kinetic proofreading [24,25] was needed to generate the affinity discrimination pattern observed in B cell affinity discrimination experiments [26]. Kinetic proofreading was simulated in an ad hoc manner, by introducing a threshold time for which antigen had to stay bound to a BCR before that BCR could engage cytoplasmic signaling molecules. The kinetic proofreading requirement, in the form of the threshold time, needed to be long enough to predominate over the competing effect of reduced serial engagement with increasing affinity, which was detrimental to affinity discrimination. In the absence of a kinetic proofreading requirement, our previous modeling studies demonstrated that the strength of B cell signaling actually decreased as affinity increased, which is the opposite of B cell affinity discrimination. However, the physical mechanism that could give rise to kinetic proofreading still needs to be explored in terms of molecular level interactions.

It is known that BCRs form oligomers in the presence of antigen [9,27], leading to the formation of the larger BCR-antigen microclusters reported in the literature. However, the molecular mechanism of BCR oligomer formation is not known. While cross-linking by soluble multivalent antigens has traditionally been used to explain BCR oligomer formation, such a mechanism cannot account for the formation of BCR oligomers in the presence of monovalent, membrane-bound antigen [7-9,22,28,29].

According to the "conformation-induced oligomerization model" of Pierce and colleagues, the force exerted by membrane-bound antigen binding to BCR within the restricted 2-D geometry of a cell-cell interaction opens up the Cμ4 domain at the base of the BCR ectodomain into a conformation that is conducive to oligomer formation [27-29]. When a BCR with an "open" Cμ4 domain encounters another BCR with its Cμ4 domain also "open", the two may form a dimer [27-29].

Here, we explored whether Pierce's confirmation-induced model of oligomer formation can account for B cell affinity discrimination. We removed the threshold time requirement, and instead introduced a requirement that only BCRs that have bound antigen can form oligomers, and only BCRs that are part of an oligomer can engage cytoplasmic signaling molecules. Such a requirement favors high affinity interactions, since as affinity increases, the antigen off-rate generally decreases. Higher affinity BCR-antigen pairs thus have a longer lifetime, which increases the odds of their encountering another BCR-antigen pair and forming an oligomer; the stability of formed oligomers is also enhanced with increasing affinity. Since only BCRs in oligomers can engage signaling molecules, there should be more signaling activity as affinity increases. Hence oligomer formation emerges as a kinetic proofreading mechanism that enables B cells to discriminate between antigens of different affinities.

## METHOD

Our method is a three-dimensional, agent-based Monte Carlo simulation of BCR-antigen binding at the surface and membrane-proximal BCR signaling in the cytoplasm. The current method is based on our previous work [26,30]. We modeled the B-cell signaling pathway as far the Src family kinase Syk. The molecular species included in our model are BCR (and its Ig-α and Ig-β signaling subunits), antigen (Ag), and the kinases Lyn and Syk. Individual molecules are explicitly simulated as discrete agents reacting with each other and diffusing subject to probabilistic parameters that can be mapped to kinetic rate constants.

*Setup*

Because we are interested in the early stages of antigen recognition, we model a single protrusion on a B cell surface, its cytoplasmic interior, and its vertical projection onto a planar bilayer containing antigen. The bilayer and B cell protrusion tip are modeled as lattices of 150×150 nodes, while the interior of the B cell protrusion is modeled to a depth of 40 nodes. We only allow one molecule per node, thus we set the spacing between nodes to 10 nm, roughly the exclusion radius of a membrane protein molecule. The domain size thus corresponds to a physical area of 1.5 μm × 1.5 μm. BCR is located on the B cell protrusion surface, antigen on the bilayer surface, Lyn is anchored below the B cell protrusion surface, and Syk is distributed in the B cell protrusion's cytoplasm. At the start of a simulation run, all of these species are distributed uniformly at random over their respective domains. At each time step, individual molecules in the population are randomly sampled to undergo either diffusion or reaction, determined by means of an unbiased coin toss.

*Reaction*

Antigen, Lyn and Syk are monovalent, while BCR molecules possess four binding sites: Two extracellular Fab domains for antigen binding, and one Ig-α and one Ig-β cytoplasmic domain, both of which serve as binding sites for Lyn and Syk. If an antigen molecule is selected for reaction, the lattice node on the B cell surface directly opposite the antigen's location is checked for a Fab domain, and if that is the case, a BCR-Ag complex may form with probability $p_{\mathrm{on(BA)}}$. If the target BCR molecule happens to also have an antigen bound on the other Fab domain, a BCR-$Ag_2$ complex will form. If a BCR molecule is selected to undergo reaction, an unbiased coin toss is performed to chose between the extracellular or cytoplasmic domains, and an additional unbiased coin toss is performed to chose one of the Fab domains (left or right), or either the Ig-α or Ig-β domain, depending on the result of the preceding coin toss. If a free Fab domain is selected, the node on the bilayer surface directly opposite is checked for antigen, which may bind with probability $p_{\mathrm{on(BA)}}$. If the selected Fab domain already has an antigen bound to it, the BCR-Ag bond may dissociate with probability $p_{\mathrm{off(BA)}}$.

If a BCR-Ag or BCR-$Ag_2$ complex is next to another BCR-Ag or BCR-$Ag_2$ complex, they may form a dimer with probability $p_{\mathrm{on(olig)}}$. This can happen either via reaction (a BCR binding an antigen right next to a BCR-Ag or BCR-$Ag_2$ complex), or diffusion (a BCR-Ag or BCR-$Ag_2$ complex moving next to another BCR-Ag or BCR-$Ag_2$ complex). Another BCR-Ag or BCR-$Ag_2$ complex may subsequently form or diffuse next to the dimer, and join the dimer with probability $p_{on(olig)}$, forming a trimer. The theoretical upper bound on the size of an oligomer is only limited by the number of BCR and antigen molecules in the simulation, whichever is smaller. All BCRs within an oligomer lose the ability to diffuse, however they gain the ability to bind Lyn on either their Ig-α or Ig-β domains. Lyn may bind with probability $p_{on(Lyn)}$, either via sampling of the Ig-α or Ig-β domains for reaction, or sampling of a Lyn molecule for reaction (similar to antigen binding on the Fab domains). Upon binding, Lyn may phosphorylate the Ig-α or Ig-β domain it is bound to with probability $p_{phos(Lyn)}$. A bound Lyn may dissociate with probability $p_{off(Lyn)}$ if the Ig-α or Ig-β domain to which it is bound is sampled. Before the dissociation trial with probability $p_{off(Lyn)}$ is carried out, a phosphorylation trial with probability $p_{phos(Lyn)}$ is performed.

If a BCR that is part of an oligomer loses all of its bound antigens through dissociation, it ceases being part of the oligomer, regaining its ability to diffuse but losing the ability to bind Lyn. Depending on whether its neighboring BCR/Ag or BCR/$Ag_2$ complexes neighbor other BCR/Ag or BCR/$Ag_2$ complexes, they may also revert to the un-oligomerized state. For example, if one of the BCRs in a dimer is no longer bound to antigen, the other BCR formerly in the dimer also regains the ability to diffuse and loses the ability to bind Lyn. Similarly, if the BCR in the middle of a trimer is no longer bound to antigen, all three BCRs revert to the un-oligomerized state. However, if the BCR at the edge of a trimer is no longer bound to antigen, the other two BCRs remain in a dimer configuration.

Syk may bind to Ig-α or Ig-β domains that have been phosphorylated by Lyn (BCRs with at least one Ig-α or Ig-β domain phosphorylated by Lyn are designated pBCR), initially with probability $p_{on(Syk)low}$. Upon binding, Syk may phosphorylate the Ig-α or Ig-β domain it is bound to with probability $p_{phos(Syk)low}$. If the phosphorylation trial by Syk is successful, the pBCR molecule with at least one Ig-α or Ig-β domain phosphorylated by Syk is designated as a ppBCR molecule. Also, if the phosphorylation trial is successful, the Syk molecule may itself become phosphorylated (phosphorylated Syk molecules are designated pSyk), either by a Lyn molecule or another pSyk molecule located on a neighboring node, with probability $p_{syk(phos)}$. If the phosphorylation trial is unsuccessful, and the pBCR molecule does not transition to the ppBCR state, the Syk molecule also remains in the inactive state (designated iSyk) and will not become phosphorylated by a Lyn or pSyk (at the start of the simulation, all Syk molecules are in the iSyk state). A phosphorylation trial of Ig-α or Ig-β by Syk occurs every time the Ig-α or Ig-β to which the Syk is attached is sampled for dissociation.

The dissociation kinetics of Syk attached to an Ig-α or Ig-β domain depend on the phosphorylation status of the Syk and the Ig-α or Ig-β domain it is bound to. If the phosphorylation trial by Syk is unsuccessful, and the Syk molecule remains in the iSyk stage and BCR molecule in the pBCR stage, the Syk molecule may dissociate with probability $p_{off(Syk)high}$. If the phosphorylation trial is successful, the Syk molecule will dissociate with probability $p_{off(Syk)high}$ from the ppBCR if the Syk molecule is successfully phosphorylated by a nearby Lyn or pSyk (becoming a pSyk molecule in the process). However, if the phosphorylation trial is successful but the Syk molecule is not phosphorylated by a nearby Lyn or pSyk molecule, the probability of the iSyk molecule dissociating is governed by $p_{off(Syk)low}$, and it remains in the iSyk state.

Later in the simulation it is possible for an iSyk molecule to encounter a ppBCR that has reached that state from previous interactions. In that case, it may bind to the Ig-α or Ig-β domains with probability $p_{on(Syk)high}$. The dissociation kinetics in such an instance are as described above: If the Syk molecule is phosphorylated by a nearby Lyn or pSyk, the probability used in subsequent dissociation trials is $p_{off(Syk)high}$ and the attached Syk molecule transitions to the pSyk state. However if the Syk molecules is not phosphorylated, it remains in the iSyk state and the probability used in subsequent dissociation trials is $p_{off(Syk)low}$.

A pSyk molecule encountering a pBCR or ppBCR may bind the Ig-α or Ig-β domains with probability $p_{on(Syk)low}$ and dissociate with probability $p_{off(Syk)high}$. In the case of a pSyk molecule binding to a pBCR molecule, the probability of the pBCR becoming phosphorylated by the pSyk molecule and transitioning to the ppBCR state is governed by $p_{phos(Syk)high}$. Lastly, a pSyk molecule that is diffusing in the cytoplasm may also phosphorylate an iSyk molecule it encounters with probability $p_{phos(SykSyk)}$. The kinetic rules for Syk are summarized in Table 1. These simulation rules for Syk are significantly more complex than those used in our previous work, and are based on a survey of the literature on the subject [31-34].

*Diffusion*

If a molecule has been selected to undergo diffusion, a random number trial with probability $p_{diff(i)}$ is used to determine whether the diffusion move will occur. The diffusion probability $p_{diff}$ is directly analogous to the diffusion coefficient $D$. The probability of diffusion of free molecules is denoted by $p_{diff(F)}$, and that of BCR-antigen complexes and BCR signalosomes by $p_{diff(C)}$. If the trial with probability $p_{diff(i)}$ is successful, a neighboring node is selected at random (4 possibilities for BCR, antigen, and Lyn, 6 possibilities for Syk) and the target node is checked for occupancy. The move will occur only if the target node is vacant, as no two molecules are allowed to occupy the same node. BCR molecules, BCR-antigen complexes, and BCR signalosomes are generally assumed to be much larger than antigen, Lyn and Syk molecules and occupy several nodes; thus in these instances there will be several target nodes that need to be vacant for the move to occur. BCR-antigen complexes and BCR signalosomes are generally assumed to diffuse slower than free molecules [22], hence $p_{diff(C)}$ is an order of magnitude lower than $p_{diff(F)}$. Since free receptor and ligand molecules are the fastest diffusing species, we set $p_{diff(F)}=1$, and $p_{diff(C)}=0.1$.

*Sampling and time step size*

In our algorithm, the entire molecular population is randomly sampled $M$ times for diffusion or reaction during every time step. Whether a diffusion or reaction trial will occur is determined by means of an unbiased coin toss. The number of trials $M$ is set equal to the total number of molecules (free plus bound) present in the system at the beginning of each time step, and the simulation is run for a number of time steps $T$.

A distinguishing feature of our method is a mapping between the probabilistic parameters of the Monte Carlo simulation and their physical counterparts. We do this by setting $p_{diff}$ of the fastest diffusing species, in this case free molecules ($p_{diff(F)}$), equal to 1 and matching that quantity to the species' measured diffusion coefficient $D$. The diffusion coefficient of free molecules on a cell membrane has been experimentally measured to be of the order of 0.1 $\mu m^2/s$ [35]. In one time step, a molecule with $p_{diff}$=1 will on average (since each molecule is on average sampled once per time step) have covered a distance of one nodal spacing, or 10 nm, giving a mean square displacement $\langle r^2 \rangle$ of $10^{-4}$ $\mu m^2$. Using $\langle r^2 \rangle = Dt$, this results in a time step size of $10^{-3}$ seconds. Once the time step size is known, it is possible to map $p_{on}$, $p_{off}$, and their ratio $P_A$ to their respective physical counterparts, $k_{on}$, $k_{off}$, and $K_A$. A detailed description of the mapping process can be found in our previous work [26]. Such a mapping makes it possible to compare our model's results to those of physical experiments to within an order of magnitude, without *a priori* setting of the simulation timescale.

<u>*Model parameters*</u>

The parameters used in our model are listed in Table 2. Parameter values used in our simulations are given on the left side of Table 2. Their physical equivalent is listed on the right side of Table 2. Where possible, the value of the physical quantity is taken from the literature, and mapped back into a value that can be used in our simulation. This is the case for BCR-antigen kinetic parameters ($P_{A(BA)}$, $p_{on(BA)}$, $p_{off(BA)}$), the number of BCR and antigen molecules ($B_0$, $A_0$), and the diffusion probability of free molecules ($p_{diff(F)}$). We vary BCR-antigen affinity by orders of magnitude across the physiological range for B cells ($K_A$=$10^5$-$10^{10}$ $M^{-1}$). BCR-antigen affinity is initially varied, as in many B cell activation experiments [7,8], by keeping $k_{on}$ constant and varying $k_{off}$. For example, in Carrasco et al. [7], affinity for the HEL series of antigen is varied by varying $k_{off}$ across five orders of magnitude, while $k_{on}$ is fixed at $2*10^6$ $M^{-1}s^{-1}$ [7,8]. This value maps to a probability of 0.1 according to our mapping scheme, while the reported $k_{off}$ values ranging from $10^1$ to $10^{-4}$ $s^{-1}$ map to $p_{off}$ values of $10^{-2}$-$10^{-7}$ for a time step size of $10^{-3}$ s. In the Supporting Information, we show the effect of varying $k_{on}$. Variation in $k_{on}$ (and/or $k_{off}$) commonly occurs when higher affinity BCRs are generated through somatic hypermutation during the affinity maturation process [36]. The literature value of $10^5$ receptors/cell [37] maps to 400 molecules for the 1.5 μm × 1.5 μm domain used in our simulations, while the antigen concentration of 10-100 molecules/$\mu m^2$ used in experiments [7] maps to 20-200 antigens. In the results shown here we used 200 antigen molecules, and results with 20 and 2000 antigen molecules are included in the Supporting Information.

We also vary the values of parameters for which we were not able to find measured values in the literature, such as the number of Lyn and Syk molecules ($L_0$, $S_0$), the probability of oligomer formation, $p_{on(olig)}$, the on and off-probabilities of cytoplasmic reactions such as $p_{on(Lyn)}$, $p_{off(Lyn)}$, $p_{on(Syk)low}$, $p_{on(Syk)high}$, $p_{off(Syk)low}$, $p_{off(Syk)high}$, and the phosphorylation reaction probabilities $p_{phos(Lyn)}$, $p_{phos(Syk)low}$, $p_{phos(Syk)high}$, $p_{Syk(phos)}$, and $p_{phos(SykSyk)}$. For the purposes of obtaining ballpark values for these parameters, we have adapted the values used in modeling studies of FcεRI-mediated signaling, which bears many similarities to BCR-mediated signaling [38,39]. Starting from the ballpark values, we performed parametric studies to gauge the effect of these parameters on affinity discrimination. These parametric studies are included as Supporting Information. Here, we show the results obtained for those parameter values that resulted in the best affinity discrimination, while still being physically reasonable. These are the values shown on the left of Table 1, with their mapped physical values shown on the right (though these are not literature values).

## RESULTS

*Histogram plots show affinity discrimination from bound antigen to phosphorylated Syk*

We investigate affinity discrimination by tabulating the quantity of several relevant molecular species at the end of a simulation run of 100 physical seconds (i.e. $10^5$ Monte Carlo simulation time steps). The quantities of interest are 1) the number antigen bound to BCR (bdAg), 2) the number of BCRs that are part of an oligomer (dBCR), 3) the number of BCRs with at least one ITAM singly phosphorylated by Lyn (pBCR), 4) the number of BCRs with at least one ITAM doubly phosphorylated by Syk (ppBCR), and 5) the number of phosphorylated Syk (pSyk). Because our simulation is stochastic in nature, these quantities vary from one run to the next. We therefore perform one thousand independent trials for each affinity value and plot the results in histograms. BCR-antigen affinity is varied by orders of magnitude across the physiological range, from $K_A$=$10^5$ $M^{-1}$ to $K_A$=$10^{10}$ $M^{-1}$, as is done in some B cell affinity discrimination experiments [7,8]. Histogram plots of the number of bound antigen molecules are shown in Figure 1. In line with experimental results [8], the number of bound antigen molecules increases with BCR-antigen affinity. The histogram plots for the four lowest affinity values ($K_A$=$10^5$ $M^{-1}$ to $K_A$=$10^8$ $M^{-1}$) are well separated, and it also is possible to distinguish between $K_A$=$10^8$ $M^{-1}$ and $K_A$=$10^9$ $M^{-1}$. The histograms for $K_A$=$10^9$ $M^{-1}$ and $K_A$=$10^{10}$ $M^{-1}$ almost completely overlap, indicative of the ceiling in affinity discrimination around $K_A$=$10^{10}$ $M^{-1}$ observed in the experimental literature [2,8].

Histogram plots of the number of dBCR are shown in Figure 2. As with bound antigen, the four lowest affinity values are clearly separated, it is possible to distinguish between $K_A$=$10^8$ $M^{-1}$ and $K_A$=$10^9$ $M^{-1}$, while the histograms for $K_A$=$10^9$ $M^{-1}$ and $K_A$=$10^{10}$ $M^{-1}$ almost completely overlap. Importantly, the number of dBCR is close to zero for all trials at the lowest affinity value, $K_A$=$10^5$ $M^{-1}$. This is in agreement with the existence of a threshold affinity for B cell activation, such as $K_A$=$10^6$ $M^{-1}$ observed in some B cell activation experiments [2,7,8], below which no B cell activation is observed. Since our simulations show that there is a nonzero number of bdAg at $K_A$=$10^5$ $M^{-1}$, this result suggests that lack of dimer formation below affinity $K_A$=$10^6$ $M^{-1}$ could be one reason for the absence of B cell activation below this affinity value.

The number of singly phosphorylated BCR, pBCR, is shown in Figure 3. The histograms are less clearly separated than in Figures 1 and 2, though it is still possible to distinguish between $K_A$=$10^7$ $M^{-1}$ and $K_A$=$10^8$ $M^{-1}$, and to a lesser extent between $K_A$=$10^8$ $M^{-1}$ and $K_A$=$10^9$ $M^{-1}$. Nevertheless, the results in Figure 3 closely resemble those of Figure 2. As in Figure 2, the number of pBCR for $K_A$=$10^5$ $M^{-1}$ is usually zero (only for a relatively few trials is this number nonzero).

The number of doubly phosphorylated BCR, ppBCR is shown in Figure 4. The pattern resembles that of Figure 3, the main difference being that the number of ppBCR for $K_A$=$10^6$ $M^{-1}$ is much smaller, and for $K_A$=$10^5$ $M^{-1}$ is zero for all trials. Thus, affinity discrimination seen in Figures 1 and 2 is maintained at the level of pBCR and ppBCR, albeit with some loss.

The number of phosphorylated Syk is shown in Figure 5. It is still possible to distinguish between $K_A$=$10^7$ $M^{-1}$, $K_A$=$10^8$ $M^{-1}$, and $K_A$=$10^9$ $M^{-1}$, thus affinity discrimination is still evident at the level of pSyk. The histograms in Figure 5 have noticeably higher standard deviations than those of Figs 1-4. Of note is that our results reproduce the threshold affinity of activation at $K_A=10^6$ $M^{-1}$ and affinity discrimination ceiling at $K_A=10^{10}$ $M^{-1}$ all the way down to the level of pSyk.

*Analysis of histogram overlaps: quantitative metric for affinity discrimination*

We quantify affinity discrimination in the histograms of Figure 1-5 using the metric Δ=(overlap area)/(m1–m2), where the area of overlap between the histograms for two adjacent affinity values is divided by m1 and m2, the histograms' mean values for those two affinities. Lower Δ values correspond to better affinity discrimination, with the best discrimination occurring at Δ=0 (no overlap between histograms). When Δ=0, it is necessary to compare the mean values of the histograms (Figure 7). In Figure 6, we show Δ for the quantities in Figures 1-5. For $10^6$ $M^{-1}/10^5$ $M^{-1}$ and $10^7$ $M^{-1}/10^6$ $M^{-1}$ there is full separation, with increases in Δ thereafter. Bound antigen has the lowest Δ, indicating a loss in affinity discrimination downstream, although this loss is proportionately small for $10^8$ $M^{-1}/10^7$ $M^{-1}$ and to a lesser extent for $10^9$ $M^{-1}/10^8$ $M^{-1}$. For $10^{10}$ $M^{-1}/10^9$ $M^{-1}$, we note that the histograms in Figures 1-5 almost completely overlap, while in Figure 6 the value of Δ is 200 or greater. We thus take Δ=200 or greater to indicate absence of affinity discrimination.

*Mean value plots show a sigmoid affinity discrimination profile with increasing affinity*

In Figure 7, we plot the trial-averaged (mean) values of the quantities represented in the histograms of Figures 1-5. Trial-averaged quantities are important as they can be thought of as analogous to the signaling response integrated from multiple protrusions on a single cell. All five quantities plotted show a monotonic, sigmoid increase with affinity, starting with a small increase from $K_A=10^5$ $M^{-1}$ to $K_A=10^6$ $M^{-1}$, a steeper increases between $K_A=10^6$ $M^{-1}$ and $K_A=10^7$ $M^{-1}$, leveling off as the ceiling of $K_A=10^{10}$ $M^{-1}$ is approached. The strength of BCR signaling is known from experiments to increase monotonically with affinity, and our simulation results reproduce this. As shown in the Supporting Information, the affinity discrimination pattern, activation threshold and affinity discrimination ceiling in our above results are generally robust with regards to variations in individual parameter values.

*Average time of initial oligomer formation shows a non-linear decrease with increasing affinity*

In Figure 8, we show the trial-averaged time of formation of the first oligomer (almost always a dimer). In every simulation trial, the time at which the first oligomer forms is recorded, and the results for each affinity value are averaged over 1000 trials and plotted in Figure 8. The average time of initial oligomer formation decreases by about half from $K_A=10^5$ $M^{-1}$ to $K_A=10^6$ $M^{-1}$, with smaller decreases as affinity increases thereafter. The relatively high time of initial oligomer formation for $K_A=10^5$ $M^{-1}$ is due to the high value of $k_{\mathrm{off}}$ at this affinity value, which results in a low rate of oligomer formation. The times shown in Figure 8 are well within the experimentally observed ~10 second timescale of BCR association with Lyn [22,23]. The standard deviation of the time of initial oligomer formation is roughly 6 s for $K_A=10^5$ $M^{-1}$, and roughly 2 s for all affinity values above that.

*Time series plots for signaling activation*

Plots of the trial-averaged number of bdAg, dBCR, pBCR, ppBCR, and aSyk as functions of time are shown in Figure 9. The number of bdAg, dBCR and pBCR approaches equilibrium after $T=10^5$ time steps, particularly for the lower affinity values, but the number of ppBCR and pSyk is still far from equilibrium. This is to be expected, as these quantities are further downstream in the signaling pathway. The plots in Figure 9 are also in agreement with the experimental finding of faster growth in signaling for higher affinity values [9]. For the lowest affinity value, $K_A=10^5$ $M^{-1}$, the number of dBCR, pBCR, ppBCR and pSyk is close to zero for all times.

## DISCUSSION

In this work, we show that the process of oligomer formation gives rise to the kinetic proofreading that we showed in previous work was necessary for B cells to discriminate antigen affinity. Our work here does not contain an explicit, ad hoc kinetic proofreading requirement, rather, kinetic proofreading emerges naturally from our simulation. Our results clearly show that just as the number of antigen bound to BCR increases with affinity, so does the number of phosphorylated Syk molecules. Though there is some loss in definition, the affinity discrimination pattern observed at the level of BCR-antigen complexes is observable at the level of phosphorylated Syk molecules.

Because of their longer lifetime, high affinity BCR-antigen pairs are more likely to encounter one another and form an oligomer. This is reflected in our finding that the average time of dimer formation decreases with affinity. More oligomer formation directly leads to more phosphorylation of BCR ITAMs by Lyn, which in turn leads to more Syk activity. If BCRs could associate with Lyn immediately upon encountering antigen, signaling strength would have decreased with affinity, as shown in our previous work [26]. The requirement that BCR not only needs to bind antigen, but also encounter another BCR-antigen pair and form a dimer, acts as a type of kinetic proofreading and results in affinity discrimination as observed in experiments. This type of oligomer formation-mediated mechanism of affinity discrimination would ensure that significant affinity discrimination is achieved even between closely related antibodies, such as those generated during the process of affinity maturation.

Our simulation results are also consistent with the existence of an experimentally observed threshold affinity of activation and affinity discrimination ceiling. Our simulations predict a B cell activation threshold at $K_A$=$10^6$ $M^{-1}$, as shown by the fact that the number of dBCR and other downstream quantities is zero below $K_A$=$10^6$ $M^{-1}$, even though the number of bound antigen at $K_A$=$10^5$ $M^{-1}$ is nonzero. This indicates that the threshold of B cell activation for downstream quantities is at least an order of magnitude higher than that for bound antigen. The precise value of the threshold affinity of activation and affinity discrimination ceiling may vary from one experimental setup to another (depending on the cell line used, antigen family, etc…), however the existence of the affinity threshold and affinity discrimination ceiling is established by experiments. Similarly, in our simulation the precise value of the threshold and ceiling may vary depending on the parameter values used, however, they are always present, as shown in the Supporting Information. The B cell activation threshold could also be modulated by lipid-mediated interactions, such as those between BCRs and raft-forming sphingolipids, which could perturb the natural lipid environment of resting B cells. For example, the formation of BCR oligomers is difficult for BCRs with low $k_{on}$, but lipid mediated mechanisms can lower the affinity threshold in such cases. A detailed model of BCR-lipid raft formation, which includes both BCR-BCR interactions (that can result in BCR oligomer formation) and BCR-lipid interactions, is currently under development.

**Table 1.** Kinetic rules for Syk used in our simulation.

| | pBCR | ppBCR |
|---|---|---|
| iSyk | Binds with $p_{on(Syk)low}$<br>Phosphorylates Ig-α/β with $p_{phos(Syk)low}$<br>Unbinds with $p_{off(Syk)high}$ | Binds with $p_{on(Syk)high}$<br>Unbinds with $p_{off(Syk)low}$ |
| pSyk | Binds with $p_{on(Syk)low}$<br>Phosphorylates Ig-α/β with $p_{phos(Syk)high}$<br>Unbinds with $p_{off(Syk)high}$ | Binds with $p_{on(Syk)high}$<br>Unbinds with $p_{off(Syk)high}$ |

**Table 2.** Parameters of our model, and their physical equivalents.

| *Simulation parameter* | *Value* | *Physical equivalent* | *Value* |
|---|---|---|---|
| $P_{A(BA)}$ | $10^{1}$-$10^{6}$ ‡ | $K_A$ BCR-antigen | $10^{5}$-$10^{10}$ $M^{-1}$ [7,8] |
| $p_{on(BA)}$ | 0.1‡ | $k_{on}$ BCR-antigen | $10^{6}$ $M^{-1}s^{-1}$ [7,8] |
| $p_{off(BA)}$ | $10^{-2}$-$10^{-7}$ ‡ | $k_{off}$ BCR-antigen | $10^{1}$-$10^{-4}$ $s^{-1}$ [7,8] |
| $B_0$ | 400 molecules | BCR molecules/cell | ~$10^{5}$ [37] |
| $A_0$ | 200 molecules ‡ | Antigen conc. | 100 molec./$\mu m^2$ [7] |
| $L_0$ | 100 molecules †‡ | Lyn molecules/cell | $2\times10^{4}$ |
| $S_0$ | 400 molecules †‡ | Syk molecules/cell | $4\times10^{5}$ |
| $p_{diff(F)}$ | 1 | $D_{\text{free molecules}}$ | 0.1 $\mu m^2/s$ [35] |
| $p_{diff(C)}$ | 0.1 | $D_{\text{complexes}}$ | ~0.01 $\mu m^2/s$ [22] |
| $p_{on(olig)}$ | 0.1‡ | $k_{on(BCR\text{-}BCR)}$ | 100 $molec^{-1}\times s^{-1}$ |
| $P_{A(Lyn)}$ | $10^{2}$ †‡ | $K_A$ Ig-α/β-Lyn | $10^{6}$ $M^{-1}$ |
| $p_{on(Lyn)}$ | 1.0 †‡ | $k_{on}$ Ig-α/β-Lyn | ~$10^{7}$ $M^{-1}$ $s^{-1}$ |
| $p_{off(Lyn)}$ | 0.01 †‡ | $k_{off}$ Ig-α/β-Lyn | ~10 $s^{-1}$ |
| $p_{on(Syk)high}$ | 1.0 †‡ | $k_{on}$ Ig-α/β-Syk | ~$10^{7}$ $M^{-1}$ $s^{-1}$ |
| $p_{on(Syk)low}$ | 0.1 †‡ | $k_{on}$ Ig-α/β-Syk | ~$10^{6}$ $M^{-1}$ $s^{-1}$ |
| $p_{off(Syk)high}$ | 1.0 †‡ | $k_{off}$ Ig-α/β-Syk | ~1000 $s^{-1}$ |
| $p_{off(Syk)low}$ | 0.001 †‡ | $k_{off}$ Ig-α/β-Syk | ~1 $s^{-1}$ |
| $p_{phos(Lyn)}$ | 0.1 †‡ | $k_{phos(Lyn)}$ | ~100 $s^{-1}$ |
| $p_{phos(Syk)high}$ | 1.0 †‡ | $k_{phos(Syk)high}$ | ~1000 $s^{-1}$ |
| $p_{phos(Syk)low}$ | 0.5 †‡ | $k_{phos(Syk)low}$ | ~500 $s^{-1}$ |
| $p_{Syk(phos)}$ | 1.0 †‡ | $k_{Syk(phos)}$ | ~1000 $s^{-1}$ |
| $p_{phos(SykSyk)}$ | 0.01 †‡ | $k_{phos(SykSyk)}$ | ~10 $s^{-1}$ |

† Represents a ballpark value calculated from [38,39]
‡ Parametric study performed

## FIGURE LEGENDS

**Figure 1. Histogram plots of the number of bound antigen molecules.** BCR-antigen affinity is varied by orders of magnitude across the physiological range in B cells, $K_A$=$10^5$ $M^{-1}$ to $K_A$=$10^{10}$ $M^{-1}$. Because of the probabilistic nature of our simulation, one thousand trials were performed for each affinity value. The parameter values used are those listed in the left column of Table 2, simulation time is $10^5$ time steps (corresponding to $T$=100 physical seconds). The number of bound antigen generally increases with affinity, as expected.

**Figure 2. Histogram plots of the number of BCRs that are part of an oligomer.** All BCRs in the simulation are checked whether they are part of an oligomer, and if they are, they are included in the tally. One thousand trials, $10^5$ time steps, with the parameters listed in the left of Table 2. As with the number of bound antigen, the number of oligomerized BCR increases with affinity.

**Figure 3. Histogram plots of the number of BCR molecules with at least one ITAM phosphorylated by Lyn.** Included in the histograms is any BCR with at least one ITAM phosphorylated by Lyn. One thousand trials, $10^5$ time steps, with the parameters listed in the left of Table 2. As with upstream quantities (bound antigen, oligomerized BCR), the number of Lyn-phosphorylated BCRs increases with affinity.

**Figure 4. Histogram plots of the number of BCR molecules with at least one ITAM phosphorylated by Syk.** Included in the histograms is any BCR with at least one ITAM phosphorylated by Syk. One thousand trials, $10^5$ time steps, with the parameters listed in the left of Table 2. As with upstream quantities, the number of Syk-phosphorylated BCR increases with affinity.

**Figure 5. Histogram plots of the number of phosphorylated Syk molecules.** Includes any Syk molecule that is in the phosphorylated state, whether attached to an Ig-α/β or freely diffusing in the cytoplasm. One thousand trials, $10^5$ time steps, with the parameters listed in the left of Table 2. As with upstream quantities, the number of twice-phosphorylated Syk increases with affinity.

**Figure 6. Quantitative comparison of affinity discrimination between adjacent affinity values.** Plots of the quantity $\Delta$=(overlap area)/($m_1$ – $m_2$), where $\Delta$, the area of overlap between the histograms for two adjacent affinity values (e.g. $K_A$=$10^7$ $M^{-1}$ and $K_A$=$10^8$ $M^{-1}$) is divided by $m_1$ and $m_2$, the histograms' mean values. A lower value of $\Delta$ corresponds to better affinity discrimination, while for the case of $\Delta$=0 one needs to compare the difference between mean values (Fig. 7). Affinity discrimination decreases with affinity, and also downstream of BCR-antigen binding.

**Figure 7. Plots of the mean values of the histogram plots in Figs. 1-5.** All five quantities plotted in the histograms of Figs. 1-5 show a monotonic, sigmoid increase with affinity.

**Figure 8. Plot of the trial-averaged time of initial dimer formation for each affinity value simulated.** For each trial, the time at which the first dimer forms is recorded, and the average over one thousand trials is plotted here. This quantity decreases with affinity, and the average time of initial dimer formation for the lowest affinity value, $K_A=10^5$ $M^{-1}$ is about twice that of higher affinity values.

**Figure 9. Plot of the mean values of the histograms in Figs. 1-5 as functions of time.** We record the mean value of the quantities plotted in Figs. 1-5 over $10^5$ time steps, and plot the results in Fig. 9. All quantities increase monotonically with affinity, with the number of bound antigen, oligomerized BCR, and Lyn-phosphorylated BCR approaching equilibrium.

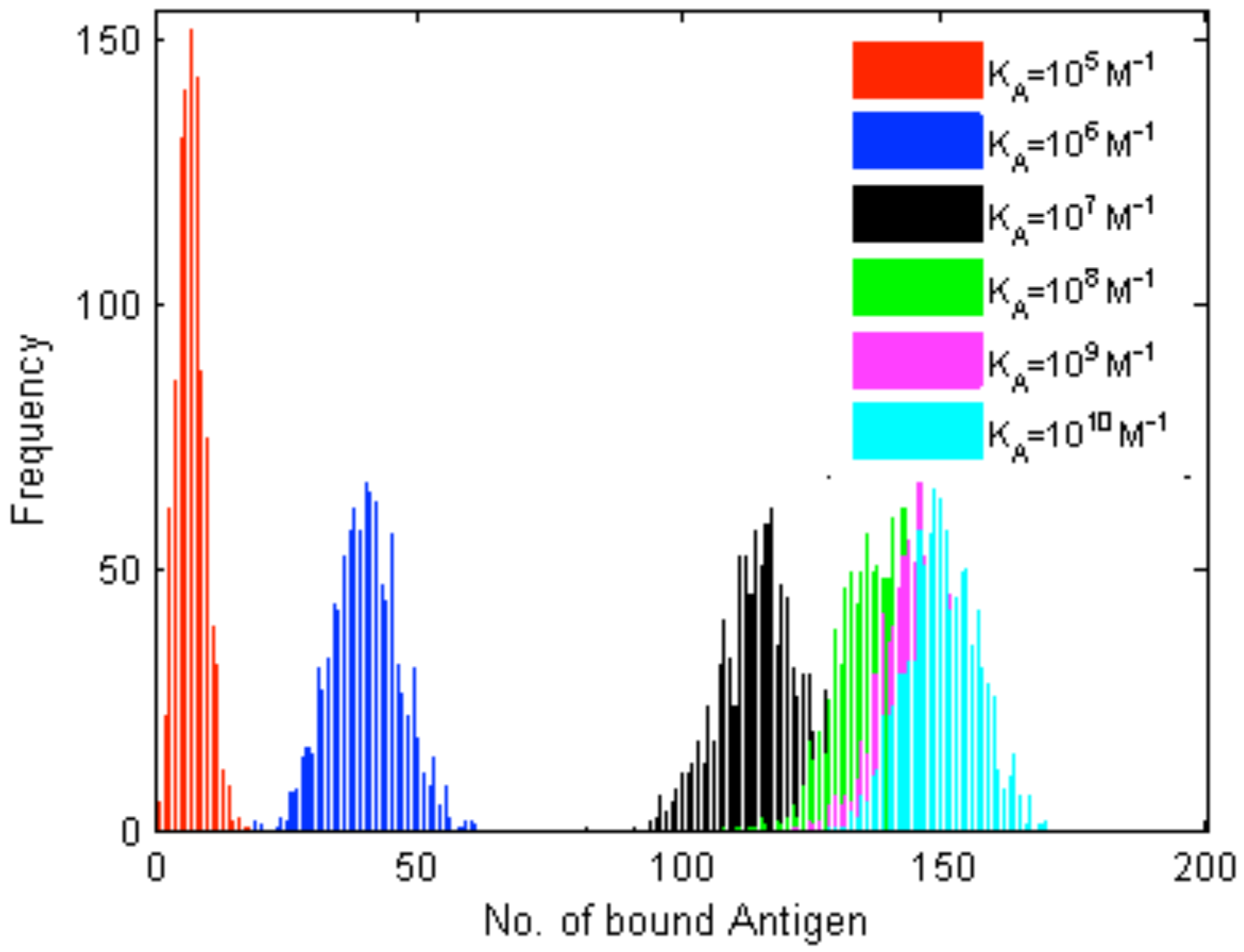

Figure 1.

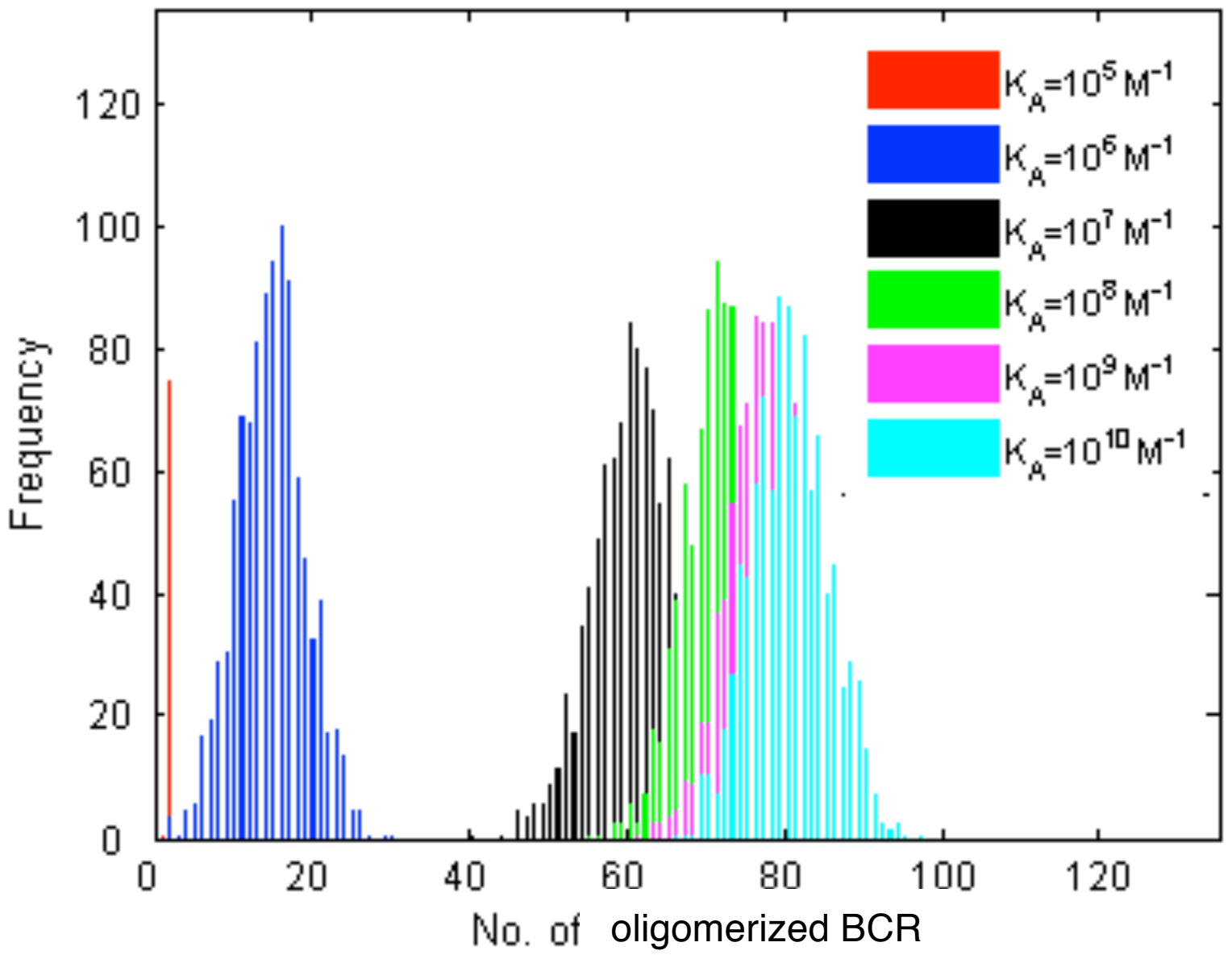

Figure 2.

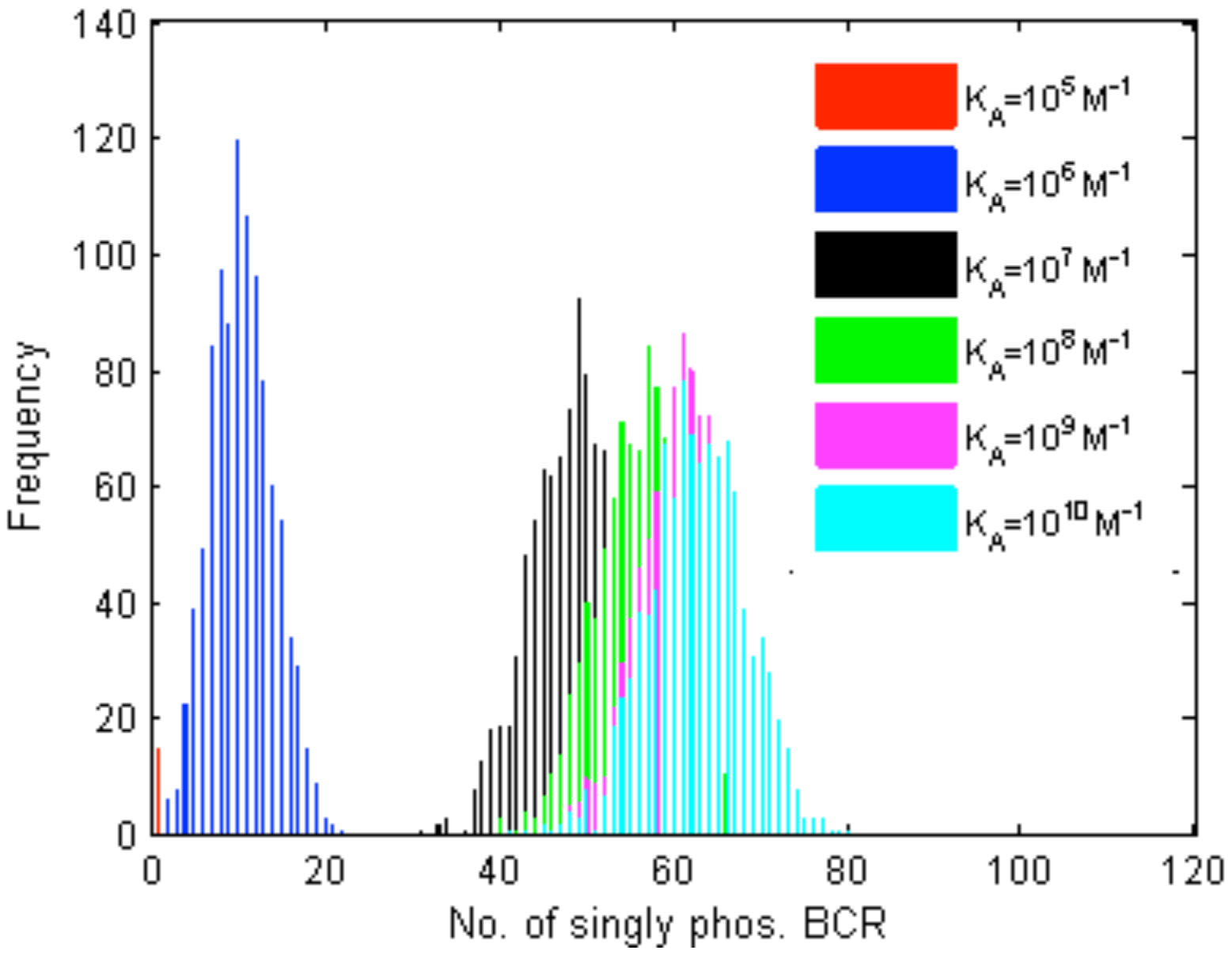

Figure 3.

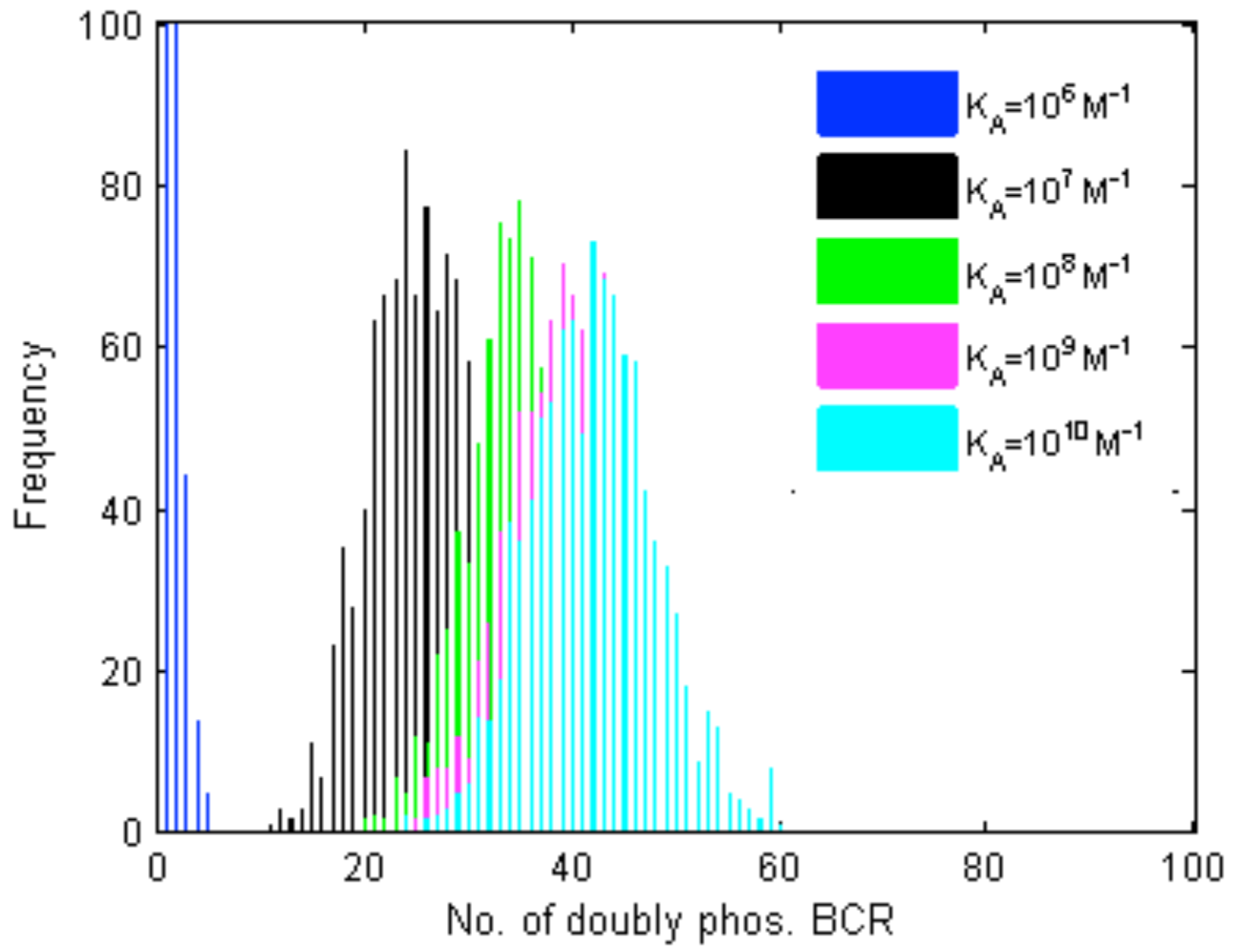

Figure 4.

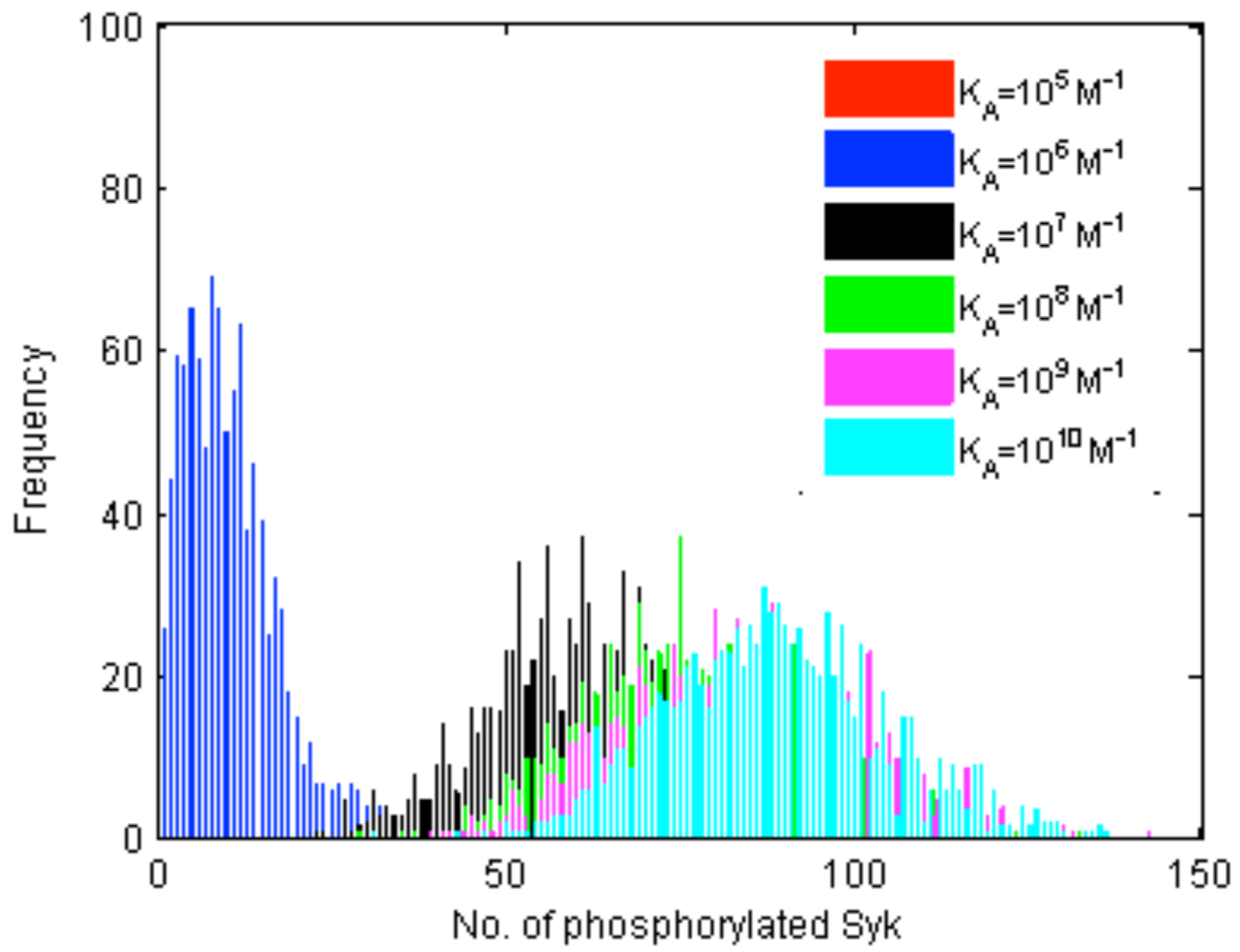

Figure 5.

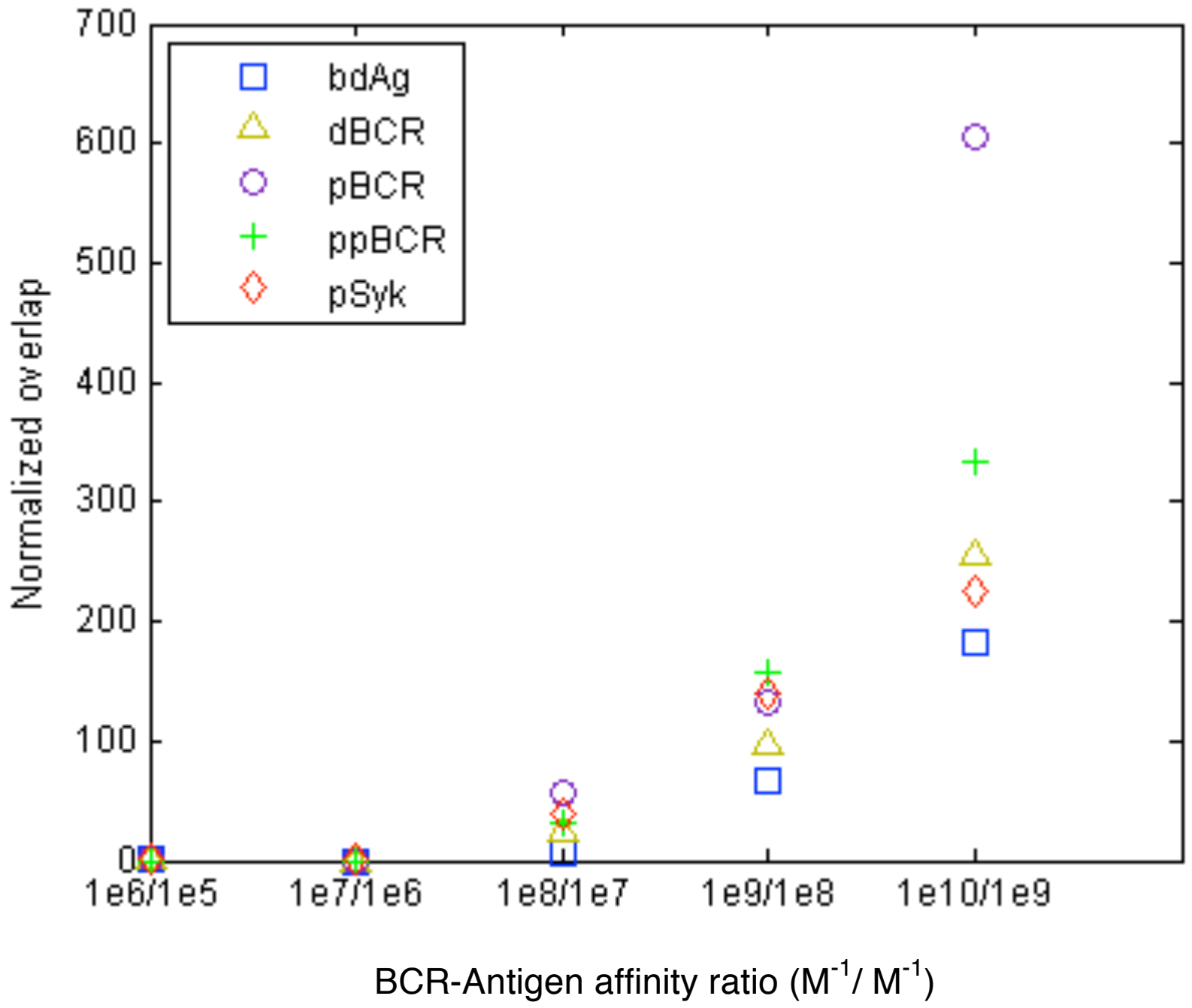

Figure 6.

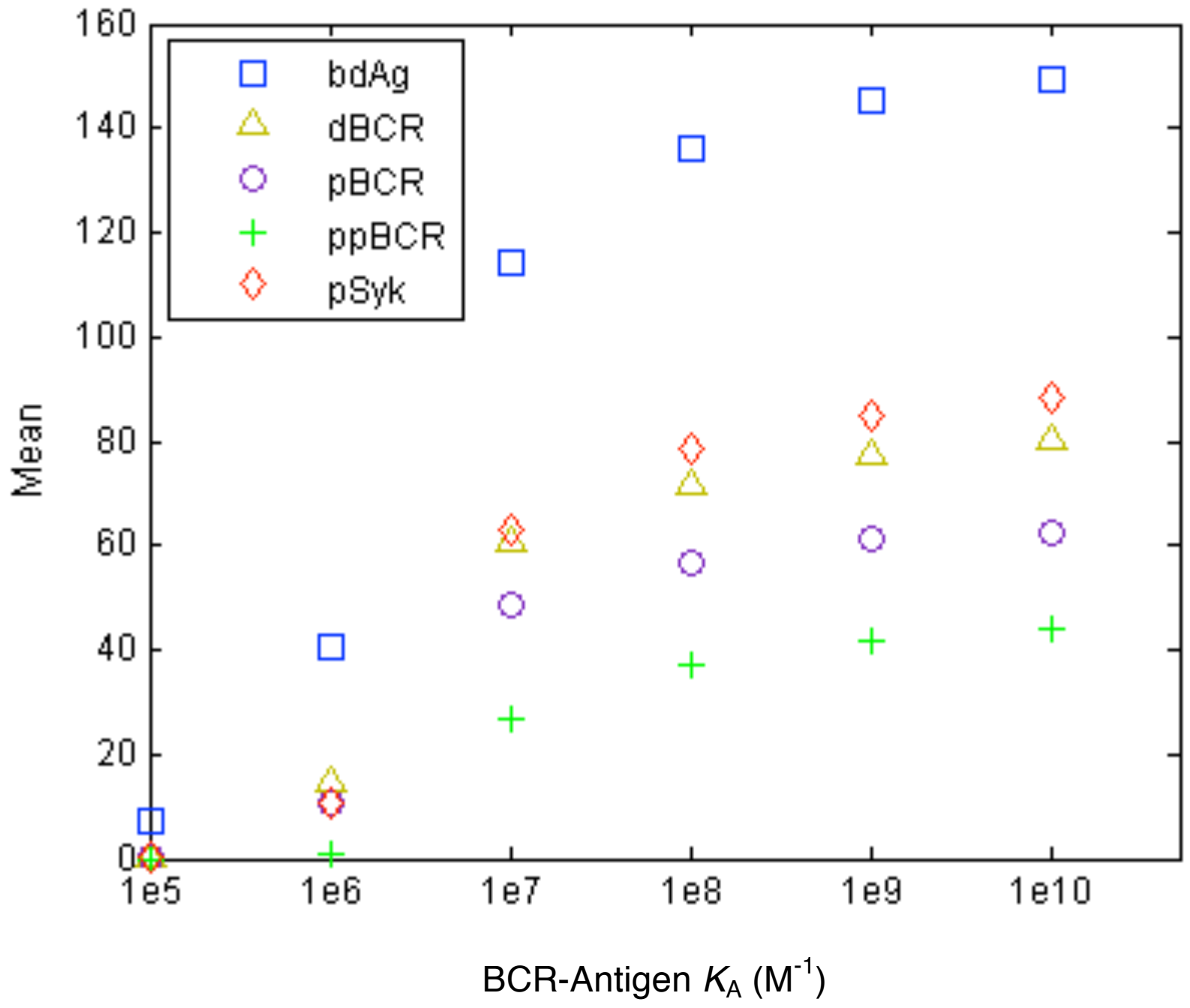

Figure 7.

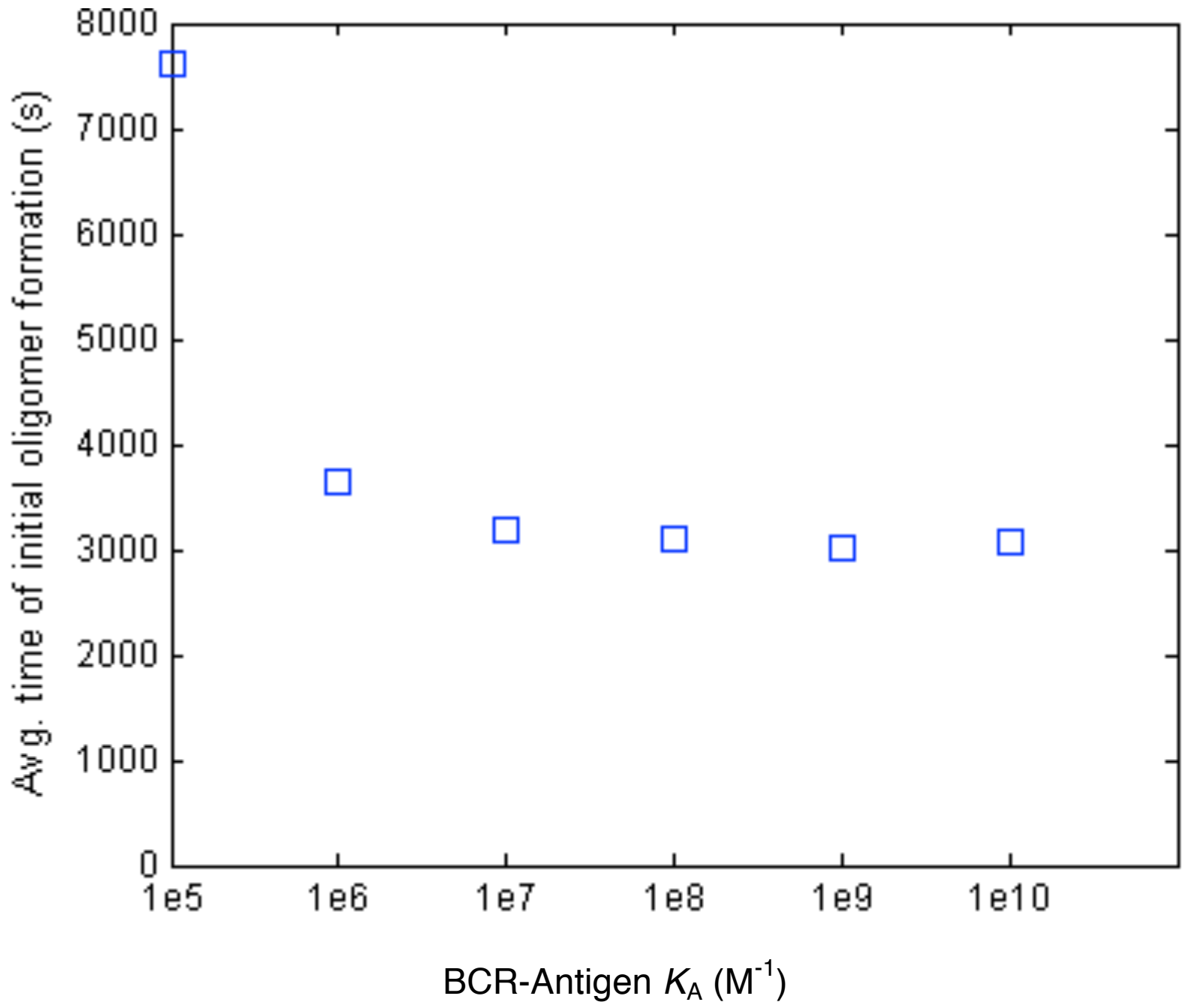

Figure 8.

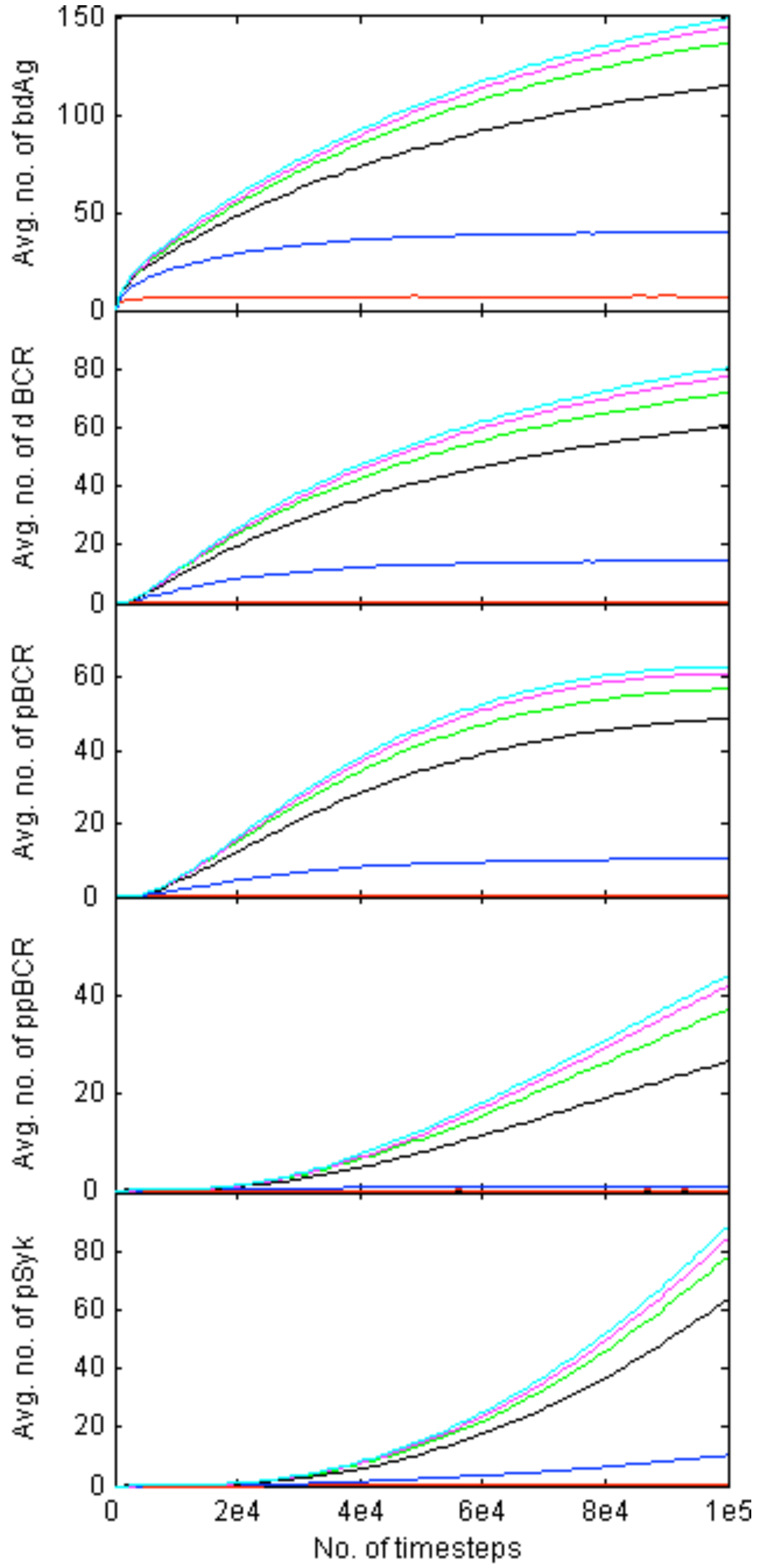

Figure 9.